\documentstyle[aps,prb,twocolumn,epsf]{revtex}
\begin{document}
\flushbottom
\draft
\twocolumn[\hsize\textwidth\columnwidth\hsize\csname
@twocolumnfalse\endcsname
\title{Excitonic ferromagnetism in the hexaborides}
\author{M. E. Zhitomirsky, T. M. Rice, and V. I. Anisimov$^*$}
\address{
Institut f\"ur Theoretische Physik,
ETH-H\"onggerberg, CH-8093 Zurich, 
Switzerland \\
$^*$Institute of Metal Physics, Ekaterinburg, Russia
}
\date{April 16, 1999}
\maketitle

\begin{abstract}
\hspace*{2mm}
A ferromagnet with a small spontaneous moment but with
a high Curie temperature can be obtained by doping an excitonic
insulator made from a spin triplet exciton condensate. Such 
a condensate can occur in 
a semimetal with a small overlap or 
a semiconductor with a small bandgap. We propose
that it is responsible for the unexpected ferromagnetism %\cite{Young}
in the doped hexaboride material Ca$_{1-x}$La$_x$B$_6$. 
\end{abstract}
%\pacs{}
\vspace*{5mm}
]
\narrowtext

In a recent letter to {\em Nature} Young {\it et al\/}. \cite{Young}
reported ferromagnetism in Ca$_{1-x}$La$_x$B$_6$ --- a material
with no partially filled $d$- or $f$-orbitals. Three features
are noteworthy, the narrow concentration range 
of doping ($0\lesssim x\lesssim 0.01$),
the small magnetic moment ($\lesssim0.07\mu_B$/La), 
and the high Curie temperature
($T_C\sim 600$~K). The parent compound CaB$_6$ is a poor conductor
which shows a small overlap $E_G$ ($<0$) between a boron-derived
valence band and a calcium-derived conduction band
at the X-point in the Brillouin zone. \cite{Massida}
The valence and conduction bands 
belong to different irreducible representations 
($X'_3$ and $X_3$, respectively)
such that the interband dipole matrix element vanishes at the X-point.
As a result the dielectric constant ($\kappa$) and the effective masses
($m_h$ and $m_e$) remain finite as $E_G\rightarrow 0$, which,
as pointed by Halperin and Rice, \cite{Halperin1}
favors formation of an excitonic condensate.

An excitonic instability for a semimetal or a semiconductor
with small values of $E_G$ was first proposed 
by Keldysh and Kopaev
\cite{Keldysh} and des Cloizeaux \cite{Cloizeaux} and later examined
in detail by a number of authors in the mid-sixties. \cite{Halperin2} 
The effective mass Hamiltonian for a gas of electrons and holes,
attracting each other through a 
Coulomb interaction screened by the background dielectric constant
$\kappa$, is treated in the Hartree-Fock approximation. 
The characteristic energy scale is set
by the exciton binding energy $E_{\rm ex}=\mu^*\kappa^{-2}$Ry, where 
$\mu^*$ is the reduced mass ($\mu^{*-1}= m_h^{-1}+m_e^{-1}$) in electron
mass units. In CaB$_6$ typical values for the effective
mass tensors are 
$m_h^\parallel = 2.17$, 
$m_h^\perp = 0.206$ (valence band) and $m_e^\parallel=0.504$, 
$m_e^\perp = 0.212$ (conduction band) from our band structure calculations
and $\kappa\sim 5$ from optical measurements 
(L. Degiorgi, private communication)
giving $E_{\rm ex}\approx 0.08$~eV. 
A triplet exciton condensate is favored over a singlet
one by the Coulomb interaction. \cite{Halperin2} In the present case
of a direct gap the broken symmetry will give rise to a local spin polarization
in each unit cell but no net magnetization at stoichiometry.
As first shown by Volkov {\it et al\/}., \cite{Volkov}
extra electrons doped into the system are 
distributed asymmetrically
between two spin projections in order to keep most
favorable pairing conditions for at least one spin species
of exciton. As a result, the excitonic insulator becomes ferromagnetic
with the Curie temperature set by the energy scale of the primary
excitonic order parameter.

The theory of a Bose condensate of
loosely bound electron-hole pairs has a close formal resemblance to 
the description of Cooper pairs in superconductors. We consider
a valence band maximum and a conduction band minimum both situated
at ${\bf k}=0$.
In the normal phase, without excitonic condensate,
the electrons states 
are classified by their wavevectors and band numbers with creation operators
$a^\dagger_{\bf k\sigma}$ and $b^\dagger_{\bf k\sigma}$ 
for valence and conduction
bands, respectively. The band quantum
number ceases to exist in the excitonic phase because of the formation 
of electron-hole pairs. This change in the groundstate is reflected
in a new Hartree-Fock average 
$\langle a^\dagger_{\bf k\sigma}b^{_{}}_{\bf k\sigma'}\rangle$,
which becomes the primary order parameter of the excitonic phase. 
Quasiparticles in two new bands are related to 
the old Bloch states via a canonical
transformation
\begin{equation}
\begin{array}{rcl}
\alpha_{\bf k\sigma} & = & u_{\bf k} a_{\bf k\sigma} +  
v_{\bf k} M_{\sigma\sigma'} b_{\bf k\sigma'} \ , \nonumber \\ 
\beta_{\bf k\sigma}  & =  & -v_{\bf k} a_{\bf k\sigma} +  
u_{\bf k} M_{\sigma\sigma'} b_{\bf k\sigma'} \ , 
\end{array}
\label{canon}
\end{equation}
where the unitary matrix $M_{\sigma\sigma'}$ describes a spin symmetry of 
the excitonic condensate. The singlet state of excitons 
corresponds to $M_{\sigma\sigma'}=\delta_{\sigma\sigma'}$, 
whereas a triplet exciton condensate 
has $M_{\sigma\sigma'}=\sum_i n_i\sigma^i_{\sigma\sigma'}$, 
$\hat\sigma^i$ being
the Pauli matrices. \cite{Halperin2}

\begin{figure}
\unitlength1cm
\epsfxsize=8.6cm
\begin{picture}(2,3)
\put(-0.4,0.3){\epsffile{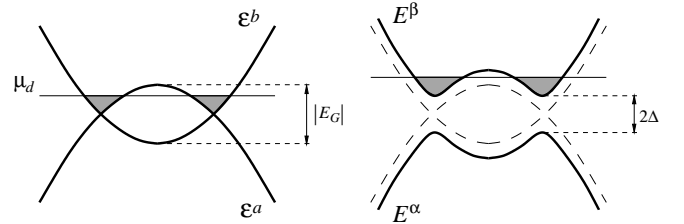}}
\end{picture}
\caption{The structure of energy spectrum in a semimetal (left)
and in an excitonic insulator (right) for one spin polarization. 
States in shaded
volumes are occupied by doped electrons.}
\end{figure}

We keep only the intraband exchange part
of the Coulomb potential between electrons and holes, 
$V({\bf k})$, which has smallest momentum transfer and 
is responsible for electron-hole binding. \cite{Halperin2}
The formation of excitons opens a gap in 
the quasiparticle spectrum $\Delta_{\bf k\sigma}= 
\sum_{\bf q}V({\bf k}-{\bf q})\langle a^\dagger_{\bf k\sigma}
b_{\bf k\sigma'}\rangle$ and the system transforms into 
an excitonic insulator see
Fig.~1. Two new bands have energy dispersions given by
$E^{\alpha,\beta}_{\bf k\sigma}=
\eta_{\bf k}\mp\sqrt{\xi^2_{\bf k}+\Delta^2_{\bf k\sigma}}$,
with $\eta_{\bf k}=
\frac{1}{2}(\varepsilon^a_{\bf k}+\varepsilon^b_{\bf k})$ 
and $\xi_{\bf k}=\frac{1}{2}(\varepsilon^a_{\bf k}-\varepsilon^b_{\bf k})$. 
The gap 
is determined from the equation
\begin{equation}
\Delta_{\bf k\sigma} = \sum_{\bf q} V({\bf k}-{\bf q}) 
\frac{\Delta_{\bf q\sigma}}{2\sqrt{\xi_{\bf q}^2 + \Delta_{\bf q\sigma}^2}}
(n^\alpha_{\bf q\sigma} - n^\beta_{\bf q\sigma}) \ ,
\label{gapE}
\end{equation}
where $n_{\bf k\sigma}^\alpha = 
\langle \alpha^\dagger_{\bf k\sigma}\alpha_{\bf k\sigma}\rangle$ 
and $n_{\bf k\sigma}^\beta= 
\langle \beta^\dagger_{\bf k\sigma}\beta_{\bf k\sigma}\rangle$ 
are occupation
numbers for $\alpha$- and $\beta$-branches of spectrum.
The neglected Coulomb terms 
with a large momentum
transfer select the triplet state with an arbitrary orientation of
polarization vector $\bf n$.\cite{Halperin2}

We start with the simplest weak-coupling variant: 
isotropic bands with masses $m_h=m_e=m$ so that
$\varepsilon^a_{\bf k} = -\varepsilon^b_{\bf k} = 
\varepsilon_0 - \frac{{\bf k}^2}{2m}-\mu$, 
$\varepsilon_0=-E_G/2$.
In the stoichiometric case the number of electrons equals 
the number of holes and $\mu=0$. For a significant overlap
there is a strong screening of the Coulomb interaction by free
carriers and we can set $V({\bf k})\approx V_0$ and, hence, 
neglect a momentum dependence of the excitonic gap.
At zero temperature the lower band is
occupied ($n^\alpha_{\bf k}\equiv 1$), 
while the upper band is empty ($n^\beta_{\bf k}\equiv 0$).
Then, the spin-independent solution of the gap equation is  
$\Delta_0 = 2\varepsilon_c \exp[-(V_0N_0)^{-1}]$, 
where $\varepsilon_c$ is a cut-off around the Fermi surface and
$N_0 = mk_F/(2\pi^2)$. The energy gain of the excitonic condensate 
over the normal phase per spin direction is
$E[0]= -\frac{1}{2} N_0\Delta_0^2$.

Electron-hole asymmetry created by doping leads to 
partial occupation of the upper $\beta$-band (Fig.~1), which 
according to Eq.~({\ref{gapE}) gradually destroys the excitonic condensate.
We relate the number of doped carries $n_d$ with a change of the
chemical potential in the normal state: $n_d = 4N_0\mu_d$. Then, 
the zero temperature excitonic gap in the unpolarized state is
\begin{equation}
\Delta_\mu^2 = \Delta_0 (\Delta_0-2|\mu_d|) \ .
\end{equation}
The excitonic condensate becomes energetically unfavorable for
$\mu_d > \Delta_0/2$.
The ground state energy 
per spin direction (relative to the stoichiometric normal state) is 
\begin{equation}
E[\mu_d] = N_0 \mu_d^2 - \frac{1}{2} 
N_0(\Delta_0-2|\mu_d|)^2 \ .
\label{Edop}
\end{equation}
Polarization is included in this picture by allowing different
chemical potentials for spin-up and spin-down quasiparticles: 
$\mu_d(1+p)$ and $\mu_d(1-p)$.
A quick check shows that the energy of a polarized excitonic insulator
$E[\mu_d(1+p)]+E[\mu_d(1-p)]$ has
a minimum value for $p=1$, i.e. for complete polarization. 
All doped electrons have parallel 
spins leaving stoichiometric conditions for carriers with opposite 
spins. This gains pairing energy relative to the unpolarized state
value $2E[\mu_d]$ and compensates for an increase in kinetic energy.
The two gaps are nonzero in the following doping ranges  
\begin{equation}
\begin{array}{rclcl}
\Delta_\downarrow & = & \sqrt{\Delta_0(\Delta_0-4\mu_d)} \ \ \ & 
 {\rm for} & 0<\mu_d<\Delta_0/4\ ;  \\ 
\Delta_\uparrow  & = & \Delta_0 & {\rm for} & 0<\mu_d<\Delta_0/2 \ .
\end{array}
\end{equation}
Note, that at intermediate concentrations electrons and holes are 
paired only for one spin direction.
At the critical point $\mu^c_d=\Delta_0/2$ the ground state energy
becomes degenerate with respect to the degree of polarization $p$
and after a jump the system returns 
back to the normal unpolarized state.

In the case of CaB$_6$ the excitonic gap, as estimated from $T_C$ 
is of the same order of magnitude or even larger
than the overlap $|E_G|=2\varepsilon_0$. 
Under this circumstances the weak-coupling approach does not work
quantitatively. In addition there is a significant 
mass anisotropy in both bands.
We, therefore, have solved the gap equation (\ref{gapE}) numerically
using effective masses from our band structure calculations and 
the gap value $\Delta_0=0.07$~eV
as suggested by optical measurements on the 5\%-doped compound 
(L. Degiorgi, private communication). 
The band overlap remains a largely unknown
parameter, which we take to be $E_G=-0.2$~eV. 
In addition, we again neglect momentum dependences of $\Delta_{\bf k}$
and $V({\bf k})$, i.e.\ assume a strong screening with a cut-off energy
$\varepsilon_c=|E_G|/2$. 
The variation of the two gaps under doping is shown
in Fig.~2. For small donor concentrations the excitonic insulator is 
again completely polarized. 
The major difference with 
the isotropic weak-coupling model is
a new concentration range, where 
the upper gap gradually decreases and 
doped electrons are only partially polarized. Note, that
ferromagnetic polarization disappears before the transition
to the normal metallic state.

\begin{figure}
\unitlength1cm
\epsfxsize=8.9cm
\begin{picture}(6,8)
\put(-0.2,0){\epsffile{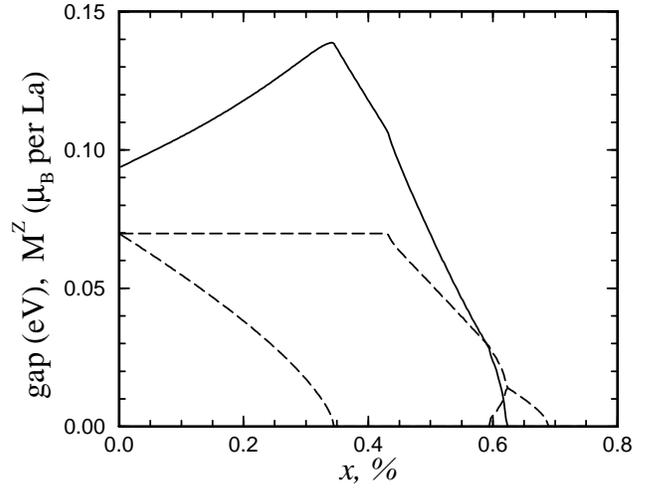}}
\end{picture}
\caption{The doping variation of the gaps for spin-up
and spin-down quasiparticles (dashed lines) and of
the saturated ferromagnetic magnetization $M^z$ (solid line)
in Ca$_{1-x}$La$_x$B$_6$ at $T=0$.}
\end{figure}

The ferromagnetic 
moment of the polarized doped excitonic insulator is 
$M^z = \sum_{\bf k\sigma} \sigma 
(n^a_{\bf k\sigma} + n^b_{\bf k\sigma})$, where
the occupation numbers are to be expressed in terms 
of the eigenstates (\ref{canon}).
The exchange electron-hole attraction does not fix
the relative orientation of the excitonic polarization vector ($\bf n$)
and the ferromagnetic moment ($M^z$). In order to determine the angle
between $\bf n$ and $\bf M$ 
we have to invoke
extra Coulomb terms.
The principal role in lifting this degeneracy
is played by a direct Hartree contribution,
which is written in terms of the Fourier harmonics of the charge
density $\rho_{\bf G}$ as $V_{\rm dir} = \frac{1}{2} \sum_{\bf G,G'}
U_{\bf G,G'}(0) \rho_{\bf G'}^* \rho^{_{}}_{\bf G}$. 
In the case of weak doping, 
we obtain from (\ref{canon}) the following expression
\begin{equation}
\rho_{\bf G}\! =\!  \sum_{\bf k\sigma} F^{ab}_{-\bf G}({\bf k},{\bf k})
u_{\bf k\sigma} v_{\bf k\sigma} (M_{\sigma\sigma}\! +\! M^*_{\sigma\sigma})
(n^\alpha_{\bf k\sigma}\! -\! n^\beta_{\bf k\sigma}) \ ,
\end{equation}
where $F^{ab}_{-\bf G}({\bf k},{\bf k})$ are the form-factors defined in
Ref.~\onlinecite{Halperin2}. In order to minimize the direct
Coulomb energy charge modulations $\rho_{\bf G}$ must vanish.
For an undoped system, when occupation numbers and coefficients
$u_{\bf k\sigma}$ and $v_{\bf k\sigma}$ are spin independent,
this requirement leads to ${\rm Tr}\hat{M}=0$, i.e.\ $V_{\rm dir}$
favors triplet states over the singlet one.
For a doped excitonic insulator, 
$n^\alpha_{\bf k\uparrow}-n^\beta_{\bf k\uparrow} \neq 
n^\alpha_{\bf k\downarrow}-n^\beta_{\bf k\downarrow}$ and
$\rho_{\bf G}$ becomes nonzero for $\hat{M}=\hat{\sigma}^z$, 
whereas it still vanishes for
$\hat{M}=\hat{\sigma}^x$ or $\hat{\sigma}^y$.
We conclude, therefore, that the perpendicular orientation
of $\bf n$ and $\bf M$ has the lowest energy.

For transverse polarization of 
triplet excitons 
an induced ferromagnetic moment is 
\begin{equation}
M^z = \sum_{\bf k\sigma} (u^2_{\bf k\sigma}-v^2_{\bf k\sigma})\sigma 
( n^\alpha_{\bf k\sigma} - n^\beta_{\bf k\sigma} )\ .
\label{smallM}
\end{equation}
In this case each of the quasiparticles (\ref{canon}) consists 
of an electron
and a hole with opposite spins and, therefore, carries only a small 
fraction of the Bohr magneton. In addition 
$M^z$ has an important contribution
from a change of the vacuum for spin-up and spin-down quasiparticles.
The contribution vanishes exactly for the constant density of states.
We used the above expression for $M^z$ to calculate the variation
of the ferromagnetic moment in CaB$_6$ under
doping, which is presented in Fig.~2.
The magnetization depends on electron-hole
asymmetry in the density of states. Smaller overlap leads generally
to more asymmetry and, consequently, to an increasing value of $M^z$.

For the above choice of parameters 
we calculated from Eq.~(\ref{gapE}) concentration dependence of
the excitonic transition temperature 
for Ca$_{1-x}$La$_x$B$_6$ and the stability boundaries
of the ferromagnetic polarization, which are shown in Fig.~3. 
The Curie temperature 
grows with doping and the maximal
$T_C$ is achieved above the concentration at
the peak value for the saturation magnetic moment.
The direct transition from
the normal semimetallic state into a polarized excitonic phase,
which exists for a certain range of doping, will 
split into two seperate transitions 
when the degeneracy lifting Coulomb 
terms are included in Eq.~(\ref{gapE}).

Excitonic ferromagnetism offers a natural explanation
of weak ferromagnetism of Ca$_{1-x}$La$_x$B$_6$.
Recognizing that the dimensionless parameter $r_S$
(ratio of interparticle spacing $r_0$ to the effective Bohr radius 
$a^*_{\rm B}$)
has a moderate value,\cite{remark} leads us to conclusion
that our
explanation is more plausible than the original interpretation
based on a ferromagnetism
of a low-density electron gas.\cite{Young,Ceperley99}
Calculations within the simplest model of an excitonic insulator
yield values of
the critical donor concentration, the weak ferromagnetic
moment, and the Curie temperature close
to the experiment. 

\begin{figure}
\unitlength1cm
\epsfxsize=8.1cm
\begin{picture}(6,8)
\put(-0.2,0){\epsffile{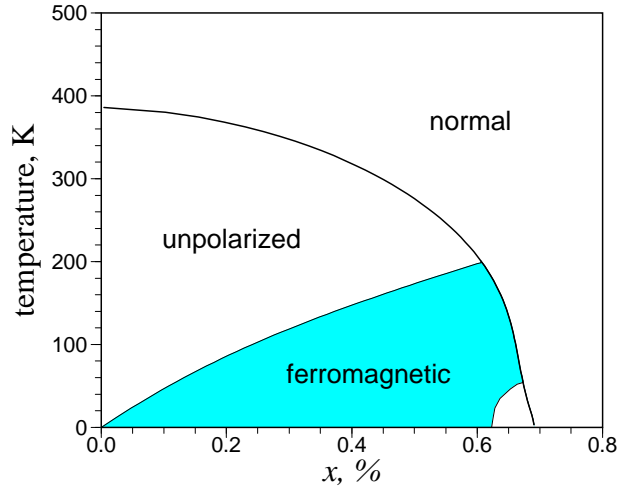}}
\end{picture}
\caption{Phase diagram of the doped excitonic insulator.
Shaded area shows stability region
of the excitonic ferromagnet.}
\end{figure}

We didn't attempt detailed theoretical predictions
mainly because of the limited experimental data available at present.
A major uncertainty is the value of the band overlap $E_G$ in 
CaB$_6$ (and in isoelectronic SrB$_6$). Our LDA calculations gave
a small overlap $E_G=-0.2$~eV, whereas the modified LDA+U method\cite{Anisimov} gave
a small band gap $E_G=0.15$~eV. In the original theory of the
excitonic insulator $E_G$ was considered to be a band parameter that
varied continuously through the value $E_G=0$.
Subsequently, the discovery of the electron-hole liquid
showed\cite{Brinkman} that a first order transition 
between states with a substantial
band gap ($E_G>0$) and overlap ($E_G<0$) should occur and that
smaller values of $|E_G|\lesssim E_{\rm ex}$ lie
in the unphysical intermediate region. At present there are
insufficient experiments to reach a firm conclusion
on this point for the hexaborides. In addition theoretical questions
such as the role of the multiple X-points in the Brillouin zone and of optical
phonons require further study.

Excitonic ferromagnetism offers a qualitative explanation for the three
key features of the ferromagnetism of Ca$_{1-x}$La$_x$B$_6$
enumerated in the beginning. Experimental predictions which will be a
decisive test of this model are optical
edges at the band gaps $2\Delta_\uparrow$ and $2\Delta_\downarrow$
in the infrared optical conductivity of doped hexaborides and also 
lowlying magnetic excitations in the stoichiometric
compounds CaB$_6$ and SrB$_6$, which are the spin-waves of the triplet
exciton condensate. 
\vspace{2mm}

{\bf Acknowledgments.} We thank L. Degiorgi and H. R. Ott for useful
discussions and comments and also L. P. Gor'kov for bringing
Ref.~\onlinecite{Volkov} to our attention.


\begin{thebibliography}{10}

\bibitem{Young}
Young, D. P. {\it et al\/}. 
High-temperature weak ferromagnetism in a low-density free-electron gas. 
{\em Nature} {\bf 397}, 412--414 (1999).  

\bibitem{Massida}
Massida, S., Continenza, A., de Pascale, T. M. \& Monnier, R. 
Electronic structure of divalent hexaborides.
{\it Z. Phys. B} {\bf 102}, 83--89 (1997).

\bibitem{Halperin1}
Halperin, B. I. \& Rice, T. M. 
Possible anomalies at a semimetal-semiconductor transition. 
{\em Rev. Mod. Phys.} {\bf 40}, 755-766 (1968).

\bibitem{Keldysh}
Keldysh, L. V. \& Kopaev, Yu. V. 
Possible instability of the semimetallic state toward coulomb 
interaction. 
{\em Sov. Phys. Solid State} {\bf 6}, 2219--2224 (1965).

\bibitem{Cloizeaux}
des Cloizeaux, J. 
Exciton instability and crystallographic anomalies in semiconductors.
{\em J. Phys. Chem. Solids} {\bf 26}, 259--266 (1965).

\bibitem{Halperin2}
For review see 
Halperin, B. I. \& Rice, T. M. 
The excitonic state at the semiconductor-semimetal transition. 
{\it in Solid State Physics} (eds. Seitz, F., Turnbull, D. \& 
Ehrenreich, H.) {\bf 21}, 115--192 (Academic Press, New York) (1968).

\bibitem{Volkov}
Volkov, B. A., Kopaev, Yu. V. \& Rusinov, A. I.
Theory of excitonic ferromagnetism.
{\em Sov. Phys. JETP} {\bf 41}, 952--959 (1975).

\bibitem{Ceperley99}
Ceperley, D. {\em Nature} {\bf 397}, 386--387 (1999). 

\bibitem{remark}
The actual value of the
parameter $r_S$, which determines
different physical regimes for electron gas, must be calculated
using the effective Bohr radius $a^*_{\rm B}$
for a Coulomb potential screened by appropriate
dielectric constant of the material. This reduces the original estimate
\cite{Young,Ceperley99} 
by almost an order of magnitude.

\bibitem{Anisimov}
Anisimov, V. I., Aryasetiawan, F. \& Lichtenstein, A. I.
First-principles calculations of the electronic structure and 
spectra of strongly correlated systems: the LDA+U method.
{\em J. Phys.: Condens. Matter} {\bf 9}, 767-808 (1997).


\bibitem{Brinkman}
Brinkman, W. F. \& Rice, T. M. 
Electron-hole liquids in semiconductors.
{\em Phys. Rev. } {\bf B7}, 1508-1523 (1973).

\end{thebibliography}
\end{document}